\documentclass[11pt]{article}

\usepackage[final]{acl}

\usepackage{times}
\usepackage{latexsym}

\usepackage[T1]{fontenc}
\usepackage[utf8]{inputenc}
\usepackage{microtype}
\usepackage{inconsolata}
\usepackage{graphicx}
\usepackage{amsmath}
\usepackage{amssymb}
\usepackage{dsfont}
\usepackage{booktabs}
\usepackage{multirow}
\usepackage{xcolor}
\usepackage{algorithm}
\usepackage{algorithmic}
\usepackage{listings}
\usepackage[most]{tcolorbox}
\usepackage{stfloats}

\definecolor{promptbg}{RGB}{248,248,248}
\definecolor{promptframe}{RGB}{180,180,180}

\newtcblisting{promptbox}{
  colback=promptbg,
  colframe=promptframe,
  boxrule=0.4pt,
  arc=2pt,
  left=4pt, right=4pt, top=4pt, bottom=4pt,
  listing only,
  listing options={
    basicstyle=\small\ttfamily,
    breaklines=true,
    breakindent=0pt,
    columns=fullflexible,
    keepspaces=true,
    aboveskip=0pt,
    belowskip=0pt,
  },
}

\newcommand{\ours}{\textsc{LoopTTS}}
\newcommand{\blfootnote}[1]{%
  \begingroup
  \renewcommand{\thefootnote}{}%
  \footnote{#1}%
  \addtocounter{footnote}{-1}%
  \endgroup
}

\title{Diagnose, Then Refine: A Closed-Loop {TTS} System with
{AudioLLM}-Guided Correction}

\author{
  \textbf{Zeyang Song\textsuperscript{1,2}},
  \textbf{Tianchi Liu\textsuperscript{3,*}},
  \textbf{Tianrui Wang\textsuperscript{4}}
  \\
  \textbf{Chenglin Xu\textsuperscript{3}},
  \textbf{Yiwen Guo\textsuperscript{5}},
  \textbf{Haizhou Li\textsuperscript{6,7,*}}
  \\[0.5em]
  \textsuperscript{1}National University of Singapore
  \quad
  \textsuperscript{2}Tencent
  \quad
  \textsuperscript{3}LIGHTSPEED
  \\
  \textsuperscript{4}Nanyang Technological University
  \quad
  \textsuperscript{5}Independent Researcher
  \\
  \textsuperscript{6}The Chinese University of Hong Kong, Shenzhen
  \quad
  \textsuperscript{7}Shenzhen Loop Area Institute
}

\begin{document}
\maketitle
\blfootnote{\textsuperscript{*}Corresponding authors.}
% ==========================================
% ABSTRACT
% ==========================================
\begin{abstract}
Current TTS systems typically rely on open-loop, single-pass generation and can produce sporadic local prosodic defects, such as misplaced stress, unnatural pauses, or flattened intonation, that utterance-level metrics often fail to expose. We present \textbf{\ours{}}, a judge-guided Filter--Judge--Refiner framework for recovering low-quality TTS outputs diagnosed by an AudioLLM. Given an initial utterance from a base TTS model, an AudioLLM Judge identifies salient prosodic issues and generates structured refine instructions; a Refiner, our fine-grained instruction-following TTS model, then performs guided expressive re-synthesis conditioned on the initial utterance, target text, and instruction. To train the Refiner, we construct \textsc{Refiner-DB}, a ${\sim}$42K-example AudioLLM-annotated dataset with word-level prosodic weak supervision. Human evaluation on diagnosed low-quality utterances shows that \ours{} can detect perceptually salient errors and correct them with the Refiner, outperforming raw generated audio and practical open-loop re-generation baselines in recovery quality. The Refiner also demonstrates stronger instruction-following ability for stress and pause control in targeted prosody modification. \ours{} is available online.\footnote{\url{https://github.com/Pooookeman/LoopTTS}}

\end{abstract}

% ==========================================
% SECTION 1: INTRODUCTION
% ==========================================
\section{Introduction}
\label{sec:intro}

% Teaser: Pipeline figure
\begin{figure*}[t]
    \centering
    \includegraphics[width=\textwidth]{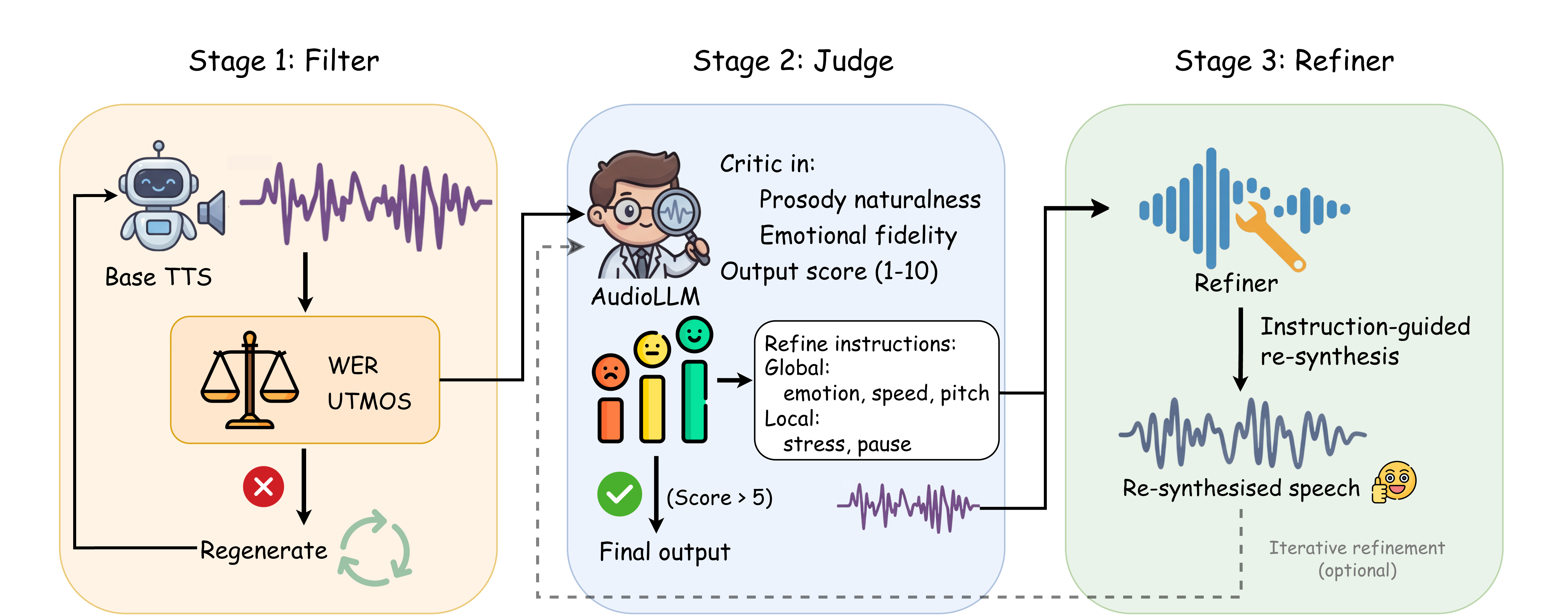}
    \caption{Overview of the \ours{} pipeline. \textbf{Stage~1 (Filter):} coarse-grained metrics (WER, UTMOS) discard catastrophically degraded outputs and trigger full re-generation. \textbf{Stage~2 (Judge):} an AudioLLM evaluates prosody naturalness and emotional fidelity, producing an overall score (1--10) along with structured refine instructions that specify both global attributes (emotion, speed, pitch) and local operations (stress, pause). Utterances meeting the threshold are accepted as final output. \textbf{Stage~3 (Refiner):} the Refiner takes the initial utterance together with the refine instructions and performs guided expressive re-synthesis, producing corrected speech with optional iterative refinement for residual defects.}
    \label{fig:overview}
\end{figure*}

The emergence of neural audio codecs has enabled a new generation of large-language-model-based text-to-speech (LLM-based TTS) systems \cite{wang2023neural,chen2024vall,du2024cosyvoice,anastassiou2024seed,chen2025f5,wang2024maskgct} that achieve near-human naturalness. Building on this foundation, controllable TTS has progressed rapidly, spanning emotion control \cite{yang2025emovoice,cho2024emosphere,liu2026emotra,wang2026evaluating} and instruction-following \cite{guo2023prompttts,yang2024instructtts,zhou2024voxinstruct,du2024cosyvoice2}, broadening the scope of speech generation.

Despite this progress, achieving consistently high-quality generation at scale remains an unsolved challenge. Most deployed TTS pipelines remain open-loop: after one generation pass, they either accept the audio or regenerate it from scratch. Due to the stochasticity of autoregressive and flow-based decoding, even strong systems can produce sporadic local prosodic defects, such as misplaced emphasis, contextually inappropriate pauses, and flattened intonation, particularly in complex linguistic contexts. These defects are subtle yet consequential; for instance, stressing a preposition (``the book IS on the table'' instead of ``the BOOK is on the table'') can make an otherwise intelligible utterance sound unnatural. Conventional objective metrics such as Word Error Rate (WER), UTMOS, and speaker similarity are useful for content, audio quality, and identity, but they operate at the utterance level and only indirectly reflect such fine-grained prosodic failures. The central question we address is therefore: \textit{how can we move beyond single-pass, open-loop TTS toward a system that integrates automatic prosodic diagnosis with bounded corrective generation?}

Existing attempts toward closing this loop remain limited. On the \textit{generation side}, the default solution for a prosodically defective utterance is full re-synthesis with a different random seed, an untargeted strategy that may fix the problem but may also reproduce it or change other attributes. Speech editing methods \cite{peng2024voicecraft,wang2024speechx} can modify existing audio, but primarily address \textit{content}-level replacement or insertion and do not expose controls such as ``add emphasis to word~X.'' On the \textit{evaluation and alignment side}, recent work has introduced AudioLLM-based judges for preference alignment \cite{zhang2024speechalign,hussain2025align2speak,xia2025mpo}; however, these approaches usually collapse diagnostic detail into scalar preference signals, discarding information needed for corrective generation, such as which word should carry stress or where a pause is missing. We therefore study whether the judge can participate in the correction loop by producing executable instructions for a separate re-synthesis model.

To address this, we present \textbf{\ours{}}, a TTS system that places an AudioLLM Judge at its core and orchestrates two functionally complementary TTS models for judge-guided corrective generation (Figure~\ref{fig:overview}). Specifically, we construct a three-stage \textbf{Filter--Judge--Refiner} architecture: a base TTS model first produces an initial utterance; the Judge identifies salient word-level prosodic issues and emits structured refine instructions; finally, the Refiner, our proposed fine-grained instruction-following TTS model, takes the initial utterance together with the refine instructions and performs guided expressive re-synthesis. This design is not waveform-local editing: unrequested regions and speaker embeddings can change, and we explicitly evaluate this trade-off. Training such a model requires word-level prosodic annotations that no existing dataset provides. We therefore leverage AudioLLM-based contrastive annotation to construct \textsc{Refiner-DB}, a ${\sim}$42K-utterance dataset with fine-grained but weak prosodic supervision. Through this judge-centric, multi-model mechanism, \ours{} studies how bounded test-time diagnosis and refinement can improve selected low-quality TTS outputs.
Our contributions are as follows:
\begin{itemize}
    \item We propose \ours{}, a judge-centric, multi-model collaborative TTS system that instantiates a closed-loop alternative through a three-stage Filter--Judge--Refiner architecture.
    \item We validate that AudioLLMs can provide scalable high-salience prosodic weak labels, and leverage this capability to construct \textsc{Refiner-DB}, a ${\sim}$42K-example dataset with word-level prosodic supervision.
    \item We introduce the Refiner, a fine-grained instruction-following TTS model trained with a position-aware objective that focuses supervision on instruction-specified prosodic positions, enabling guided expressive re-synthesis of flagged prosodic defects in our LoopTTS.
\end{itemize}

% ==========================================
% SECTION 2: RELATED WORK
% ==========================================
\section{Related Work}
\label{sec:related}

\paragraph{Controllable and expressive TTS.}
Recent LLM-based TTS systems \cite{chen2024vall,chen2025f5,wang2024maskgct, koeltts, stickertts, Glmtts, llasa, indextts2} have established a dominant backbone for high-quality speech synthesis. Built on these backbones, controllable TTS~\cite{controllable_survey} has progressed along several axes: emotion control~\cite{cho2024emosphere, 10832181, yang2025emovoice, EMORLTTS}, natural-language description~\cite{guo2023prompttts, yang2024instructtts}, and instruction-guided synthesis~\cite{zhou2024voxinstruct, du2024cosyvoice2}. In parallel, speech editing methods \cite{peng2024voicecraft,wang2024speechx} enable post-hoc modification of existing audio by replacing or inserting words at the content level. However, none of these methods can accept prosodic-level refine instructions to refine an already-generated utterance.

\paragraph{LLM-as-Judge for TTS quality optimization.}
AudioLLMs are increasingly used to improve TTS quality. On the evaluation side, recent LLM-as-a-Judge models \cite{zheng2023judging} such as AudioJudge \cite{manakul2025audiojudge} and SpeechJudge \cite{zhang2025speechjudge} can perceive fine-grained acoustic details and prosody to assess speech naturalness, approaching human-level judgment. Because human preference data is expensive to collect and inter-annotator agreement is often low, a growing body of work adopts these LLM judges to provide preference signals for TTS post-training: RLAIF-SPA \cite{yang2025rlaif} uses LLM judgments of semantic accuracy and prosody-emotion alignment as RLAIF rewards, and Step-Audio-EditX \cite{yan2025step} applies reinforcement learning with LLM-derived rewards for expressive speech editing. Agent-based frameworks such as DialogueAgents \cite{dialogueagents} modify text or sentence-level emotional expression through multi-agent coordination. However, these methods do not focus on detecting prosodic defects or performing fine-grained refinement of an already-generated utterance.

% ==========================================
% SECTION 3: METHOD (Pipeline + Data only)
% ==========================================
\section{The \ours{} Framework}
\label{sec:framework}

This section presents the \ours{} framework: a three-stage inference pipeline for iterative refinement (Section~\ref{ssec:pipeline}) and a contrastive data construction paradigm for training the Refiner (Section~\ref{ssec:data_pipeline}). The AudioLLM capabilities underlying both are validated in Section~\ref{sec:validation}.

\subsection{Three-Stage Quality Assurance Pipeline}
\label{ssec:pipeline}

The pipeline is designed as a cascade that operates on utterances produced by a base TTS model, where each stage narrows the set of utterances requiring expensive processing (Figure~\ref{fig:overview}): coarse filtering removes catastrophic failures at negligible cost, AudioLLM diagnosis evaluates the remainder, and the Refiner re-synthesizes only the flagged subset. This makes \ours{} an offline quality-assurance pipeline rather than a streaming TTS model.

\paragraph{Stage~1: Coarse-grained filtering.} Given target text $T$, the base TTS model generates $y_\text{init}$. Stage~1 removes severe content or quality failures using WER and UTMOS \cite{saeki2022utmos}: an utterance passes only when Whisper-large-v3-turbo yields WER${\leq}3.0\%$ against $T$ and UTMOS${>}3.0$. Otherwise, Stage~1 performs up to two additional re-generation attempts with different random seeds. Utterances that still fail are excluded from later stages; the rest proceed to Stage~2.

\paragraph{Stage~2: AudioLLM prosodic diagnosis.} For each retained utterance, an AudioLLM Judge scores prosodic naturalness and emotional fidelity on a 1--10 scale. Utterances scoring at least $\tau_\text{judge}{=}5$ are accepted as-is; lower-scoring utterances are flagged and assigned a refine instruction $I$ covering global style (emotion, speed, pitch) and local prosody (stress, pause). The concrete Judge choices are validated in Section~\ref{sec:validation} and specified in the experimental setup. Full prompts are provided in Appendix~\ref{app:prompts}.

\paragraph{Stage~3: Guided expressive re-synthesis.} The Refiner re-synthesizes flagged utterances conditioned on the initial audio, refine instruction, and target text:
$y_R = \text{Refiner}(y_\text{init},\, I,\, T)$. This is guided expressive re-synthesis rather than waveform-level local editing, so unrequested regions and speaker embeddings can change; we therefore evaluate speaker similarity and the resulting expressiveness--similarity trade-off. The output can be fed back to Stage~2 for bounded iterative refinement.

\subsection{Contrastive Instruction Data Construction}
\label{ssec:data_pipeline}
% Data construction figure
\begin{figure*}[t]
    \centering
    \includegraphics[width=\textwidth]{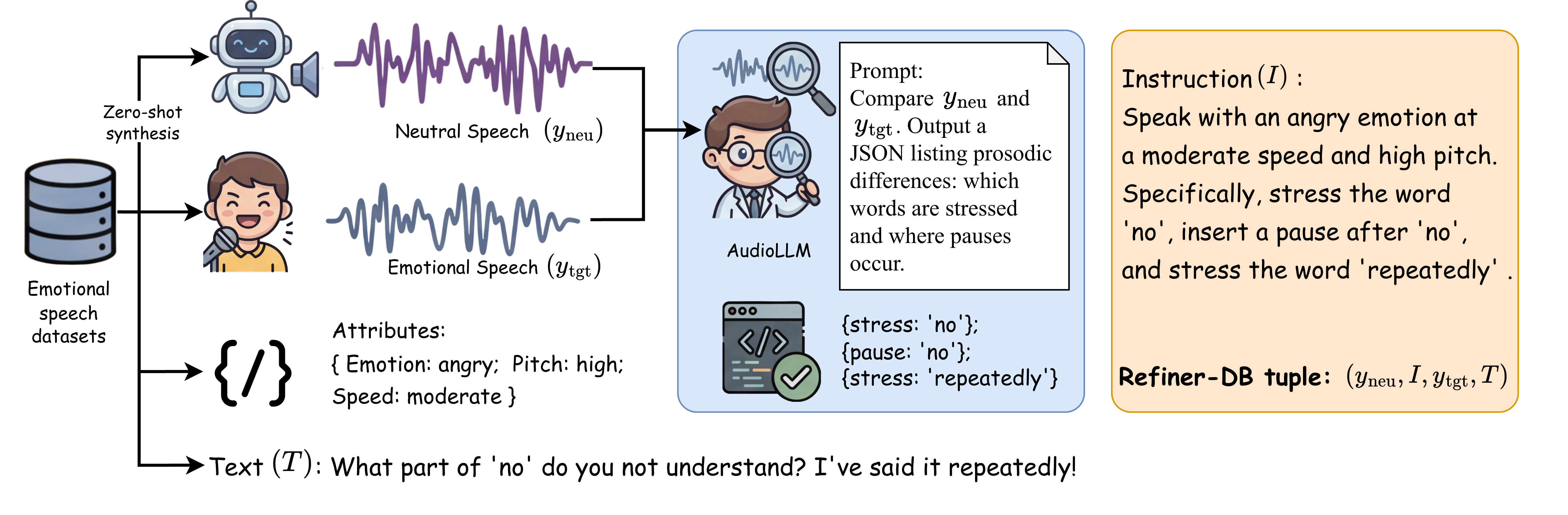}
    \caption{Contrastive data construction for \textsc{Refiner-DB}. Each expressive recording $y_\text{tgt}$ is paired with a zero-shot neutral clone $y_\text{neu}$ synthesized from the same text and speaker with neutral emotion. Global style comes from dataset metadata; an AudioLLM compares the two utterances and outputs word-level prosodic differences (stress, pause) as structured JSON. Both sources are merged into a natural-language refine instruction $I$, yielding tuples $(y_\text{neu}, I, y_\text{tgt}, T)$. Full prompts are in Appendix~\ref{app:prompts}.}
    \label{fig:data_construction}
    % \vspace{-0.10 in}
\end{figure*}

Training the Refiner requires word-level prosodic supervision, which is expensive to label manually. We therefore construct \textsc{Refiner-DB} through contrastive AudioLLM annotation (Figure~\ref{fig:data_construction}): each expressive recording $y_\text{tgt}$ is paired with a neutral synthetic counterpart $y_\text{neu}$ of the same text and speaker, and the AudioLLM extracts salient high-contrast prosodic differences between them. Global style metadata and local AudioLLM annotations are merged into a refine instruction $I$, yielding training tuples $\mathcal{D} = \{(y_\text{neu}, I, y_\text{tgt}, T)\}$.

This procedure yields ${\sim}$42K weakly annotated contrastive tuples. At inference time, $y_\text{init}$ plays the same role as an utterance to be improved. Dataset composition and filtering details are provided in Appendix~\ref{app:data_stats}, and full AudioLLM prompts are provided in Appendix~\ref{app:prompts}.

% ==========================================
% SECTION 4: MODEL AND TRAINING
% ==========================================
\section{Refiner: Architecture and Training}
\label{sec:model}
\label{ssec:ce_loss}

\begin{figure*}[t]
    \centering
    \includegraphics[width=\textwidth]{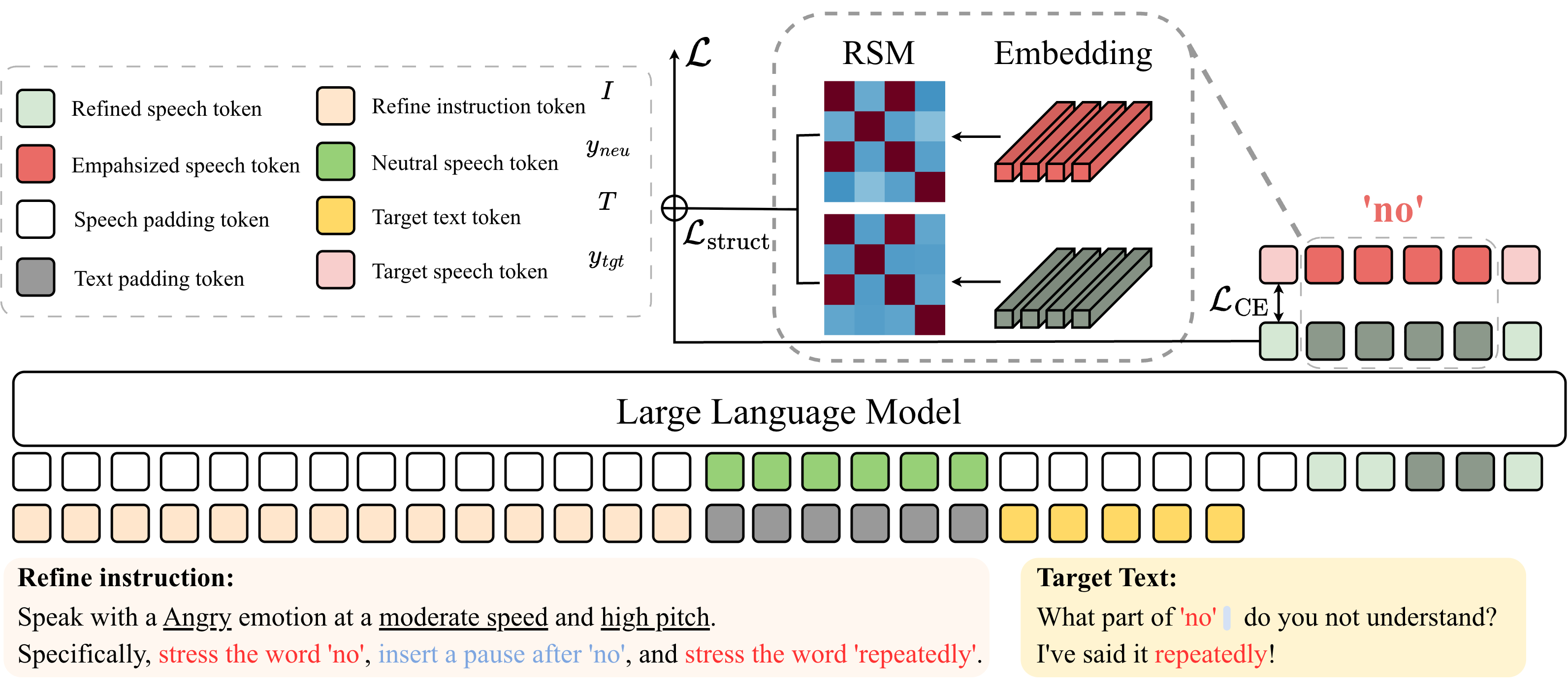}
    \caption{Refiner architecture and position-aware training objective. The Refiner takes a refine instruction $I$, the initial utterance $y_\text{init}$, and the target text $T$ as input. Two losses are applied in training: a position-weighted CE loss $\mathcal{L}^{\text{CE}}$ that amplifies gradients at stressed and pause positions, and a structural alignment loss $\mathcal{L}_{\text{struct}}$ that aligns span-level consistency.}
    \label{fig:refiner_diagram}
    \vspace{0.10 in}
\end{figure*}

The Refiner must execute refine instructions from the AudioLLM Judge, yet standard training objectives lack the inductive bias to focus on the sparse, prosodically critical positions that carry the refine signal. In a typical utterance, only 10--30 out of 200+ codec tokens correspond to instruction-specified stress or pause positions, making it difficult for uniform objectives to supervise these regions. We address this with a position-aware training objective (Figure~\ref{fig:refiner_diagram}): token-level loss reweighting that amplifies gradients at instruction-specified positions (Section~\ref{ssec:ce_loss}), and span-level relational alignment that encourages internal coherence within stress spans (Section~\ref{ssec:structural_loss}).

\subsection{Architecture}
\label{ssec:architecture}

Our Refiner is built on EmoVoice \cite{yang2025emovoice}, a codec language model for emotion-controllable TTS that autoregressively predicts 50\,Hz CosyVoice \cite{du2024cosyvoice} semantic tokens, decoded into waveforms via flow matching and HiFi-GAN \cite{kong2020hifi}. We initialize from the pretrained checkpoint and fine-tune on \textsc{Refiner-DB}. The input sequence is the concatenation $(I,\; y_\text{init},\; T)$: placing the refine instruction $I$ first ensures that it conditions all subsequent processing under causal attention; the initial utterance $y_\text{init}$ precedes the target text $T$ to provide a concrete acoustic reference.

\subsection{Position-Weighted Cross-Entropy Loss}

Under standard cross-entropy (CE) loss, all tokens contribute equally to the gradient, leaving the learning signal at prosodically critical positions diluted by the majority of unchanged tokens. Our Position-Weighted CE loss assigns higher weights to instruction-specified stress and pause positions, forcing the model to focus on these sparse but decisive regions.

\paragraph{Prosody span identification.} We use Whisper word-level timestamps\footnote{\url{https://github.com/linto-ai/whisper-timestamped}} to locate the codec token spans corresponding to each stressed word and each pause position specified in the refine instruction. These prosodic spans, i.e., the regions where the Refiner must deviate from the input, serve both the Position-Weighted CE (all spans) and the Structural Alignment Loss (stress spans only; Section~\ref{ssec:structural_loss}). We encode them as a per-token position label $m_t \in \{0, 1, 2\}$:
\[
m_t = \begin{cases}
1 & \text{if position } t \text{ falls within a stress span,} \\
2 & \text{if position } t \text{ falls within a pause span,} \\
0 & \text{otherwise.}
\end{cases}
\]

\paragraph{Position weighting.} Each token receives a weight determined by its prosodic role:
\begin{equation}
    w_t = 1 + \alpha \cdot \mathds{1}[m_t{=}1] + \beta \cdot \mathds{1}[m_t{=}2]
    \label{eq:weight}
\end{equation}
Stress and pause tokens thus receive mild upweighting relative to the unit baseline, encouraging the model to attend to the annotated local edits without overwhelming global speech modeling. The Position-Weighted CE loss is:
\begin{equation}
    \mathcal{L}^{\text{CE}} = \frac{\sum_t \text{CE}(\hat{y}_{t},\; y_{t}) \cdot w_t \cdot v_t}{\sum_t w_t \cdot v_t}
    \label{eq:ce_loss}
\end{equation}
where $v_t = \mathds{1}[y_{t} \neq \text{pad}]$ masks padding positions. In practice we set $\alpha {=} 0.2$ and $\beta {=} 0.3$, a gentle boost that prioritizes prosodic positions without destabilizing the generation of other attributes (emotion, pitch, speed).

% Refiner diagram
\subsection{Position-Aware Structural Alignment Loss}
\label{ssec:structural_loss}

The Position-Weighted CE optimizes each token independently, but stress realization is inherently span-level, spanning 5--20 contiguous tokens. Token-wise optimization may predict each position correctly in isolation yet fail to capture the internal structure of a stress span. To encourage span-level consistency, we introduce a structural alignment loss based on Representational Similarity Matrices (RSMs) \cite{park2019relational,kornblith2019similarity}, aligning pairwise token relations within each stress span between prediction and target. (Pause spans are excluded because inserted pauses consist almost entirely of silence tokens, whose trivial internal structure does not benefit from relational alignment.)

\paragraph{RSM construction and alignment.} Using the prosodic spans identified in Section~\ref{ssec:ce_loss}, let $\mathcal{P} = \{\mathcal{S}_1, \mathcal{S}_2, \ldots\}$ denote the set of all stress spans (maximal contiguous regions where $m_t = 1$). For each span $\mathcal{S} = [s, e)$, let $\mathbf{H}_\mathcal{S} \in \mathbb{R}^{|\mathcal{S}| \times D}$ denote the acoustic-code embeddings within the span. After row-wise $\ell_2$ normalization ($\bar{\mathbf{H}}_\mathcal{S}$), the RSM is:
\begin{equation}
    \mathbf{R}_\mathcal{S} = \bar{\mathbf{H}}_\mathcal{S} \, \bar{\mathbf{H}}_\mathcal{S}^\top \in \mathbb{R}^{|\mathcal{S}| \times |\mathcal{S}|}
    \label{eq:rsm}
\end{equation}
where each entry is the cosine similarity between two positions in the span. During teacher-forced training, we construct $\mathbf{R}^{\text{pred}}_\mathcal{S}$ from the model hidden representations at acoustic-code positions and $\mathbf{R}^{\text{tgt}}_\mathcal{S}$ from the corresponding target acoustic-code embeddings, making the loss an embedding-level alignment objective. We then minimize the squared Frobenius distance averaged over all spans:
\begin{equation}
    \mathcal{L}_{\text{struct}} = \frac{1}{|\mathcal{P}|} \sum_{\mathcal{S} \in \mathcal{P}} \frac{1}{|\mathcal{S}|^2} \left\| \mathbf{R}^{\text{pred}}_\mathcal{S} - \mathbf{R}^{\text{tgt}}_\mathcal{S} \right\|_F^2
    \label{eq:struct_loss}
\end{equation}
Dividing by $|\mathcal{S}|^2$, the number of entries in the RSM, ensures scale invariance across spans of different lengths. When $\mathcal{P} = \emptyset$, the loss is set to zero.

The overall training objective combines both terms:
\begin{equation}
    \mathcal{L} = \mathcal{L}^{\text{CE}} + \lambda \cdot \mathcal{L}_{\text{struct}}
    \label{eq:total_loss}
\end{equation}
We set $\lambda {=} 0.1$ in all experiments.

% ==========================================
% SECTION 5: EXPERIMENTS
% ==========================================
\section{AudioLLM Evaluation}
\label{sec:validation}

Before deploying AudioLLMs in our pipeline, we validate two capabilities: (1)~\textit{diagnosing}, whether their inference-time diagnoses align with human perception (Section~\ref{ssec:validation_diagnostic}), and (2)~\textit{annotation}, whether they can extract salient prosodic labels for training data construction (Section~\ref{ssec:validation_annotation}). We benchmark five AudioLLMs on evaluation data held out from the training set, sourced from EmoVoice (Emotional) and LibriTTS (Regular). Detailed evaluation setups are provided in Appendix~\ref{app:eval_setup}.

\subsection{Diagnostic Capability}
\label{ssec:validation_diagnostic}

Each AudioLLM diagnoses prosodic defects in 100 TTS-generated utterances (50 Regular, 50 Emotional) and generates refine instructions in the same format as Stage~2 (prompt in Appendix~\ref{app:prompts}). Six non-technical evaluators with fluent English listening proficiency then rate whether each diagnosis aligns with their own judgment on a 1--4 scale (4: fully aligned, 3: mostly aligned, 2: partially aligned, 1: not aligned).

% Diagnostic quality table
\begin{table}[t]
\centering
\small
\resizebox{\columnwidth}{!}{
\begin{tabular}{lcc}
\toprule
\textbf{AudioLLM} & \textbf{Regular} & \textbf{Emotional} \\
\midrule
Gemini-3-flash \cite{team2023gemini}       & 3.1 & 3.2 \\
Gemini-3-pro \cite{team2023gemini}        & \textbf{3.3} & \textbf{3.4} \\
Kimi-Audio \cite{kimiteam2025kimiaudiotechnicalreport}  & 2.2 & 2.5 \\
Qwen3-Omni \cite{Qwen3-Omni}    & 2.6 & 2.7 \\
Qwen3-Omni Thinking \cite{Qwen3-Omni}       & 2.9 & 3.0 \\
\bottomrule
\end{tabular}
}
\caption{Mean human--AudioLLM alignment scores (1--4). Higher is better.}
\label{tab:diagnostic_quality}
\vspace{0.10 in}
\end{table}

\begin{table}[t]
\centering
\small
\resizebox{\columnwidth}{!}{
\begin{tabular}{lcccc}
\toprule
 & \multicolumn{2}{c}{\textbf{Stress}} & \multicolumn{2}{c}{\textbf{Pause}} \\
\cmidrule(lr){2-3} \cmidrule(lr){4-5}
\multirow[c]{-2}{*}{\textbf{AudioLLM}} & \textbf{Rec.} & \textbf{F1} & \textbf{Rec.} & \textbf{F1} \\
\midrule
Gemini-3-flash       & 0.861 & 0.535 & 0.573 & 0.432 \\
Gemini-3-pro         & \textbf{0.884} & \textbf{0.673} & \textbf{0.697} & \textbf{0.532} \\
Kimi-Audio           & 0.588 & 0.438 & 0.440 & 0.316 \\
Qwen3-Omni           & 0.613 & 0.441 & 0.487 & 0.330 \\
Qwen3-Omni Thinking  & 0.667 & 0.465 & 0.478 & 0.325 \\
\bottomrule
\end{tabular}
}
\caption{AudioLLM annotation quality (micro Recall and F1) against professional human labels. The labels mark salient stress and pause differences rather than exhaustive prosodic structure.}
\label{tab:annotation_quality}
\vspace{0.10 in}
\end{table}

Results (Table~\ref{tab:diagnostic_quality}) show that Gemini-3-pro achieves the highest alignment among the evaluated AudioLLMs on both subsets (3.3 Regular, 3.4 Emotional). We therefore use it as the main Judge, while treating its diagnoses as fallible signals rather than ground truth.

\subsection{Annotation Capability}
\label{ssec:validation_annotation}

We assess whether AudioLLMs can identify salient prosodic differences between neutral and expressive speech, the core capability required for constructing \textsc{Refiner-DB}. Professional annotators label prominent stress and pause positions on 200 neutral--expressive utterance pairs; each AudioLLM independently annotates the same pairs (prompt in Appendix~\ref{app:prompts}) and is evaluated against these labels using micro recall and F1. This is a weak-labeling evaluation: the goal is to obtain scalable high-salience supervision, not exhaustive word-level prosody annotation.

Results (Table~\ref{tab:annotation_quality}) show that Gemini-3-pro achieves the highest recall and F1 on both stress and pause, making it the best annotator among the tested AudioLLMs for the large-margin contrasts used in our data construction pipeline (Section~\ref{ssec:data_pipeline}). The absolute F1 values remain moderate, especially for pause (0.532), so we do not interpret the labels as precise or exhaustive word-level ground truth. We adopt Gemini-3-pro for \textsc{Refiner-DB} because it provides useful scalable weak supervision, and we later evaluate final outputs with human listeners rather than using these labels alone.

% ==========================================
\section{Experiments}
\label{sec:experiments}

\subsection{Metrics}
\label{ssec:setup}

We evaluate the content consistency of synthesized speech using WER, with transcriptions obtained from Whisper-large-v3-turbo \cite{radford2023robust}. Speaker identity preservation is measured by speaker similarity (SIM), the cosine similarity between WavLM-based speaker embeddings of the output and the reference audio\footnote{\url{https://github.com/BytedanceSpeech/seed-tts-eval}}. We also report UTMOS \cite{saeki2022utmos} as an automatic speech quality estimator. For MOS-style subjective evaluation, ten professional paid evaluators with fluent English listening proficiency independently score each sample in a blind shuffled test. MOS rates overall naturalness and audio quality (1--5), while MOS-I rates instruction-following fidelity across five dimensions (emotion, pitch, speed, stress, pause), each on a 1--5 scale. When per-evaluator means are available, subjective scores are reported with 95\% confidence intervals (subscript $\pm$) computed from the $t$-distribution. For paired evaluation, the same evaluator pool performs blind A/B comparisons over 30 flagged samples per comparison; we report the preferred system, preference rate, and the number of samples with at least 7/10 and 9/10 evaluator agreement. Scoring rubrics and reliability statistics are provided in Appendices~\ref{app:scoring} and~\ref{app:human_reliability}.

\subsection{LoopTTS Pipeline Evaluation}
\label{ssec:pipeline_eval}

We evaluate the full \ours{} pipeline on CosyVoice2-generated speech. We sample 100 Stage~2 flagged utterances (score ${<}\,5$) per condition (Neutral and Emotional, 200 total) for human evaluation. We choose this selected failure subset for two reasons: only a small fraction of the full set is routed to Stage~3 refinement, so full-distribution MOS would be dominated by already accepted utterances and make recovery gains harder to distinguish; and full-distribution MOS over all systems would require substantially more human ratings. Appendix~\ref{app:objective_eval} reports full-pipeline stage accounting and objective metrics over 3{,}000 utterances.

\begin{figure}[t]
\centering
\includegraphics[width=\columnwidth]{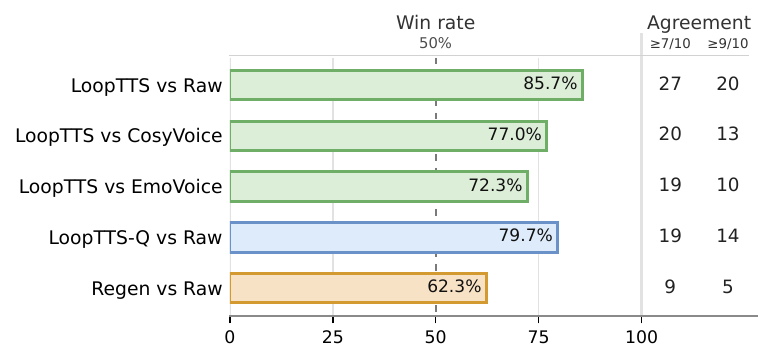}
\caption{Blind paired preference evaluation on flagged utterances. Each paired comparison uses 30 samples.}
\label{fig:paired_preference}
\vspace{0.10 in}
\end{figure}

Table~\ref{tab:pipeline_eval} compares raw flagged utterances, budget-aligned no-target re-generation, and closed-loop refinement. Re-generation uses the same initial output and Stage~2 Judge trigger as \ours{}, but spends the one-correction budget on untargeted full re-synthesis rather than the instruction-conditioned Refiner. The Global-only control retains emotion, speed, and pitch instructions while removing word-level stress and pause cues. \ours{}-Q keeps the Gemini-annotated Refiner unchanged and replaces the entire inference-time Judge with Qwen3-Omni Thinking \cite{Qwen3-Omni}, including both diagnosis and instruction generation. Figure~\ref{fig:paired_preference} reports win rate for the first-listed system and agreement at 7/10 and 9/10 listeners.

% Pipeline evaluation table
\begin{table*}[t]
\centering
\small
\begin{tabular*}{\textwidth}{l @{\extracolsep{\fill}} ccc ccc}
\toprule
 & \multicolumn{3}{c}{\textbf{Neutral}} & \multicolumn{3}{c}{\textbf{Emotional}} \\
\cmidrule(lr){2-4} \cmidrule(lr){5-7}
\multirow[c]{-2}{*}{\textbf{Condition}} & \textbf{WER}$\downarrow$ & \textbf{SIM}$\uparrow$ & \textbf{MOS}$\uparrow$ & \textbf{WER}$\downarrow$ & \textbf{SIM}$\uparrow$ & \textbf{MOS}$\uparrow$ \\
\midrule
\multicolumn{7}{l}{\textit{Baseline}} \\
\quad Before refinement       & 2.04 & \textbf{0.73} & 3.21$_{\pm.15}$ & 2.11 & 0.69 & 3.01$_{\pm.16}$ \\
\midrule
\multicolumn{7}{l}{\textit{Re-generation (open-loop)}} \\
\quad CosyVoice2 \cite{du2024cosyvoice2}              & 1.95 & 0.71 & 3.45$_{\pm.14}$ & 2.03 & \textbf{0.70} & 3.15$_{\pm.15}$ \\
\quad EmoVoice \cite{yang2025emovoice}                & 1.83 & 0.68 & 3.90$_{\pm.13}$ & \textbf{1.72} & 0.65 & 3.85$_{\pm.12}$ \\
\midrule
\multicolumn{7}{l}{\textbf{\ours{}} \textit{(closed-loop)}} \\
\quad \ours{} (Global-only)  & 1.87 & 0.68 & 3.96$_{\pm.10}$ & 1.82 & 0.66 & 3.72$_{\pm.16}$ \\
\quad \ours{} (1-round)     & 1.86 & 0.68 & 4.17$_{\pm.12}$ & 1.79 & 0.67 & 4.01$_{\pm.14}$ \\
\quad \ours{}-Q (1-round)     & 1.90 & 0.67 & 4.09$_{\pm.16}$ & 1.84 & 0.66 & 3.98$_{\pm.15}$ \\
\quad \ours{} (2-round)     & 1.85 & 0.66 & 4.16$_{\pm.13}$ & 1.80 & 0.64 & 4.17$_{\pm.13}$ \\
\quad Human-instruction       & \textbf{1.79} & 0.67 & \textbf{4.20}$_{\pm.14}$ & 1.78 & 0.67 & \textbf{4.26}$_{\pm.12}$ \\
\bottomrule
\end{tabular*}
\caption{Pipeline recovery evaluation on Stage~2 flagged utterances under Neutral and Emotional conditions. \ours{} uses Gemini-3-pro as the inference-time Judge, while \ours{}-Q uses the same Refiner with the inference-time Judge changed to Qwen3-Omni Thinking.}
\label{tab:pipeline_eval}
\vspace{0.10 in}
\end{table*}

% Main comparison table
\begin{table*}[t]
\centering
\small
\resizebox{\textwidth}{!}{
\begin{tabular}{lccc ccccc c}
\toprule
 & \multicolumn{3}{c}{\textbf{Objective}} & \multicolumn{6}{c}{\textbf{Subjective}} \\
\cmidrule(lr){2-4} \cmidrule(lr){5-10}
\multirow[c]{2}{*}{\textbf{Model}} & \multirow[c]{2}{*}{\textbf{WER}$\downarrow$} & \multirow[c]{2}{*}{\textbf{SIM}$\uparrow$} & \multirow[c]{2}{*}{\textbf{UTMOS}$\uparrow$} & \multirow[c]{2}{*}{\textbf{MOS}$\uparrow$} & \multicolumn{5}{c}{\textbf{MOS-I}$\uparrow$} \\
\cmidrule(lr){6-10}
 & & & & & \textbf{Emo.} & \textbf{Pitch} & \textbf{Speed} & \textbf{Stress} & \textbf{Pause} \\
\midrule
CosyVoice2$^\dagger$       & -- & \textbf{0.72} & 3.40 & 2.21$_{\pm.11}$ & 2.25$_{\pm.12}$ & 2.31$_{\pm.10}$ & 2.54$_{\pm.09}$ & 2.15$_{\pm.10}$ & 2.12$_{\pm.09}$ \\
CosyVoice2-marker       & 4.74 & 0.69 & 3.27 & 2.95$_{\pm.16}$ & 3.02$_{\pm.11}$ & 3.78$_{\pm.12}$ & 3.46$_{\pm.09}$ & 4.32$_{\pm.07}$ & 3.88$_{\pm.08}$ \\
EmoVoice         & \textbf{2.81} & 0.67 & \textbf{3.86} & \textbf{4.32}$_{\pm.13}$ & \textbf{4.13}$_{\pm.11}$ & 3.45$_{\pm.09}$ & 3.76$_{\pm.08}$ & 3.83$_{\pm.08}$ & 2.92$_{\pm.09}$ \\
\textbf{Refiner}          & 2.90 & 0.68 & 3.60 & 4.21$_{\pm.14}$ & 4.04$_{\pm.10}$ & \textbf{4.01}$_{\pm.10}$ & \textbf{4.13}$_{\pm.08}$ & \textbf{4.57}$_{\pm.07}$ & \textbf{4.28}$_{\pm.08}$ \\
\bottomrule
\end{tabular}
}
\caption{Instruction-following comparison. $^\dagger$CosyVoice2 frequently reads the instruction text verbatim, making its WER not comparable.}
\label{tab:main_results}
% \vspace{0.10 in}
\end{table*}

% Loss ablation table
\begin{table}[t]
\centering
\small
\resizebox{\columnwidth}{!}{
\begin{tabular}{lccc}
\toprule
\textbf{Variant} & \textbf{WER}$\downarrow$ & \textbf{MOS}$\uparrow$ & \textbf{Avg.\ MOS-I}$\uparrow$ \\
\midrule
\textbf{Refiner}       & 2.90 & \textbf{4.21}$_{\pm.13}$ & \textbf{4.21}$_{\pm.09}$ \\
w/o $\mathcal{L}_\text{struct}$          & \textbf{2.78} & 4.06$_{\pm.15}$ & 4.01$_{\pm.09}$ \\
w/o Position-Weighting        & 2.92  & 4.17$_{\pm.14}$ & 4.13$_{\pm.08}$ \\
w/o Both       & 2.87 & 4.03$_{\pm.14}$ & 3.75$_{\pm.10}$ \\
\bottomrule
\end{tabular}
}
\caption{Ablation: effect of loss components.}
\label{tab:ablation_loss}
% \vspace{-0.10 in}
\end{table}

Table~\ref{tab:pipeline_eval} shows that a single \ours{} refinement round improves MOS over both raw flagged audio and the strongest re-generation baseline under Neutral and Emotional conditions. Removing local stress and pause cues reduces MOS from 4.17/4.01 to 3.96/3.72 while WER and SIM remain comparable, supporting the contribution of position-specific instructions beyond global style control. Figure~\ref{fig:paired_preference} supports the recovery result: listeners prefer \ours{} over raw flagged audio (85.67\%), CosyVoice2 re-generation (77.00\%), and EmoVoice re-generation (72.3\%). For the latter comparison, 19/30 samples receive at least 7/10 listener agreement and 10/30 receive at least 9/10 agreement. Because the re-generation baselines use the same judge-triggered one-correction budget, these results support targeted guided expressive re-synthesis over a no-target retry strategy.

The held-out Judge result further shows that the framework is not tied to Gemini at inference time. \ours{}-Q replaces Gemini-3-pro with Qwen3-Omni Thinking for both diagnosis and instruction generation, yet obtains MOS scores of 4.09/3.98 under Neutral/Emotional conditions and remains preferred over raw flagged audio in paired evaluation (79.67\%). Because our Refiner performs guided expressive re-synthesis rather than local waveform editing, SIM reduction is an expected trade-off when prosody or emotion is strengthened. Similar trade-offs have also been observed in expressive TTS studies \cite{cocoemo2026,diemo_tts2025}. 

\subsection{Instruction-Following Comparison}
\label{ssec:main_results}

This experiment isolates the Refiner's instruction-following capability outside the full cascade. Human evaluators use MOS-I to judge whether each requested control--emotion, pitch, speed, stress, or pause--is executed naturally without introducing harmful unintended changes. We compare against CosyVoice2 \cite{du2024cosyvoice2}, CosyVoice2-marker, and EmoVoice \cite{yang2025emovoice}. All systems receive the same instruction $I$ and text $T$; only the Refiner additionally conditions on the initial utterance $y_\text{init}$. Appendix~\ref{app:baselines} provides the baseline setup details and limitations.

Results are shown in Table~\ref{tab:main_results}. The Refiner achieves the highest MOS-I on Pitch, Speed, Stress, and Pause under this protocol, indicating that fine-tuning on \textsc{Refiner-DB} improves execution of the annotated prosodic controls. The slight MOS gap relative to EmoVoice is largely due to occasional less vivid emotional delivery, suggesting that the Refiner is better suited as a Stage~3 corrective model than as a standalone emotional TTS generator.

\subsection{Ablation Studies}
\label{ssec:ablation}

% \paragraph{Is audio conditioning necessary?} The Refiner conditions on the original audio $y_\text{init}$, which provides an acoustic prior for selective preservation. Removing $y_\text{init}$ reduces the model to standard instruction-following TTS, directly testing whether conditioning on existing speech is what enables targeted prosodic modification rather than generation from scratch. Results are shown in Table~\ref{tab:ablation_audio}.
%
% % Audio conditioning ablation table
% \begin{table}[t]
% \centering
% \small
% \begin{tabular}{lcccc}
% \toprule
% \textbf{Variant} & \textbf{WER}$\downarrow$ & \textbf{SIM}$\uparrow$ & \textbf{UTMOS}$\uparrow$ & \textbf{MOS}$\uparrow$ \\
% \midrule
% \ours{}           & -- & -- & -- & -- \\
% w/o $y_\text{init}$  & -- & -- & -- & -- \\
% \bottomrule
% \end{tabular}
% \caption{Ablation: effect of conditioning on the original audio. Removing $y_\text{init}$ reduces the model to standard instruction-following TTS.}
% \label{tab:ablation_audio}
% \end{table}
We ablate Position-Weighting and Structural Alignment  to isolate their individual contributions.

Results (Table~\ref{tab:ablation_loss}) suggest that the two loss components help instruction-following, but the evidence should be interpreted cautiously. Removing $\mathcal{L}_\text{struct}$ causes the largest Avg.\ MOS-I drop, consistent with stress realization benefiting from span-level relational alignment. Removing position weighting produces a smaller but consistent MOS/MOS-I decrease, suggesting that token-level emphasis on annotated spans is useful. However, WER is slightly better for some ablated variants and the absolute MOS gaps are modest, so these ablations support the losses as helpful design choices rather than uniquely isolating the mechanism behind all gains. The separate audio-conditioning ablation in Appendix~\ref{app:audio_conditioning_ablation} shows that removing $y_\text{init}$ only slightly affects MOS but reduces SIM and MOS-I, indicating that the initial utterance helps preserve speaker/prosodic identity and avoid unnecessary edits.

Notably, the full Refiner exhibits a minor WER increase compared to ablated variants. This reflects an inherent trade-off: more aggressive prosodic modification, particularly emphatic stress and inserted pauses, occasionally causes higher WER (e.g., Whisper misrecognizes elongated stressed syllables or transcribes deliberate pauses as punctuation).

% ==========================================
% SECTION 6: CONCLUSION
% ==========================================
\section{Conclusion}
\label{sec:conclusion}

We have presented \ours{}, an offline Filter--Judge--Refiner framework that integrates prosodic diagnosis with guided expressive re-synthesis. The key idea is to leverage AudioLLMs as scalable but imperfect sources of diagnostic weak supervision and pair them with a Refiner trained on contrastive prosody data and a position-aware objective. Experiments show that the Refiner improves fine-grained instruction following under our protocol, and that bounded refinement improves human-rated recovery quality on Stage~2 flagged utterances compared with raw audio and practical open-loop re-generation baselines. The modular architecture allows the inference-time Judge to be replaced, as shown by \ours{}-Q, although the training labels remain Gemini-derived. Extending the framework to full-distribution human evaluation, stronger deployment baselines, tonal languages, and lower-cost open judges are important directions for future work.

% ==========================================
% SECTION 7: LIMITATIONS
% ==========================================
\section*{Limitations}
\label{sec:limitations}

The main human pipeline evaluation is conducted on Stage~2 flagged utterances, directly testing recovery from diagnosed defects. This subset is where refinement can change the output; on the full distribution, the observed effect would be diluted by many utterances that already pass without refinement. Due to the cost of blind human listening tests, we do not run full-distribution human evaluation over every initial-generation utterance. Appendix~\ref{app:objective_eval} instead reports larger-scale full-pipeline accounting and objective metrics over 3{,}000 utterances.

Our experiments focus on English read and emotional speech with a limited range of speakers and corpora. The framework is modular in that the initial TTS system and inference-time Judge can be replaced, as illustrated by the re-generation baselines and \ours{}-Q, but broader speaker, language, and speaking-style coverage remains future work. Extending the framework to larger multilingual, multi-speaker emotional corpora and incorporating prompt-based speaker conditioning are natural next steps.

Current annotations and controls represent stress and pause decisions as binary labels. This supports word-level diagnosis and correction, but does not model degrees of stress intensity or calibrated pause duration. Future extensions could use more fine-grained prosodic labels to support continuous or multi-level expressive control.

The Refiner performs guided expressive re-synthesis conditioned on the original audio rather than waveform-level local editing. This design supports broad prosodic correction, but expressive or prosodic changes may also shift speaker embeddings. Such SIM degradation is a common trade-off in expressive TTS, where emotional/prosodic changes can affect speaker similarity \cite{cocoemo2026,diemo_tts2025}; Appendix~\ref{app:speaker_similarity_tradeoff} discusses this trade-off in more detail.

% ==========================================
% ETHICS STATEMENT
% ==========================================
\section*{Ethics Statement}
\label{sec:ethics}

\textbf{Human annotation and fair compensation.}
Human participation in this work includes ten professional paid blind listening test evaluators with fluent English listening proficiency, six diagnostic alignment raters, and professional prosodic annotators who provided annotation ground-truth labels and human-instruction upper-bound references.
All participants were compensated in accordance with local labor regulations and institutional guidelines, consistent with ACL requirements regarding fair treatment and remuneration of human participants.
All human tasks were limited to listening to AI-synthesized speech and providing subjective quality ratings; no personally identifiable information was collected from any participant. Our institution classifies such perception-only studies as minimal-risk research, exempt from formal ethics board review.

\textbf{Data privacy and consent.}
All training data in \textsc{Refiner-DB} are constructed from publicly available speech corpora released for research purposes (ESD, RAVDESS, SAVEE, MESS, EmoVoice-DB, and LibriTTS).
The contrastive instruction data are generated entirely from these public datasets via our AudioLLM-annotated data construction pipeline; no private, user-uploaded, or personally identifiable data are used at any stage.
For any dataset release, we will follow the license and redistribution terms of each source corpus: derived annotations and construction scripts will be released where permitted, while source audio will be redistributed only when the original license allows it.
The released dataset and model checkpoints do not contain any private user data.

\textbf{Licensing and responsible use.}
The complete code, data construction scripts, prompts, model checkpoints, generated annotations, raw AudioLLM JSON outputs where license-compatible, and evaluation sample lists will be released upon acceptance for non-commercial academic research.
We acknowledge that controllable prosodic speech synthesis carries potential risks of misuse, including generating deceptive or manipulative audio content.
We emphasize that \ours{} is intended as a research contribution to advance fine-grained prosodic control in TTS, and we encourage responsible use with appropriate human oversight in any downstream application.

\textbf{Usage of AI assistants.}
AI language models were used in this work in three capacities: (1)~language polishing during paper writing, (2)~prosodic annotation during dataset construction (Appendix~\ref{app:annotation_prompt}), and (3)~prosodic diagnosis in the Judge stage of the \ours{} pipeline (Appendix~\ref{app:diagnostic_prompt}). All experimental design, analysis, and scientific conclusions were made by the authors.

% ==========================================
% REFERENCES
% ==========================================
\section*{Acknowledgments}

This work was supported by the Program for Guangdong Introducing
Innovative and Entrepreneurial Teams (Grant No.\ 2023ZT10X044).

\bibliography{custom}

% ==========================================
% APPENDIX
% ==========================================
\appendix

\section{LoopTTS on scaled evaluation set}
\label{app:objective_eval}

Beyond the small-scale human evaluation in Section~\ref{ssec:pipeline_eval}, we conduct a larger-scale objective comparison on 3{,}000 emotional synthesis utterances (Table~\ref{tab:objective_eval}). Each sample provides a neutral reference audio and a target emotion label; the system must synthesize emotionally expressive speech for the given text. We report WER~($\downarrow$), SIM~($\uparrow$), and UTMOS~($\uparrow$). The high error rate of CosyVoice2 and EmoVoice is largely due to generation failures (e.g., garbled audio, missing words); Stage~1 performs up to two additional re-generation attempts after the initial generation and removes most such failures before Judge-based diagnosis. Samples that still fail the Stage~1 filters after these attempts are not passed to Stage~2, while the remaining samples enter optional refinement.

To contextualize full-distribution performance, we additionally apply untargeted CosyVoice2 and EmoVoice re-generation directly to the same 3{,}000 synthesis requests. We also evaluate a best-of-5 alternative: for each request, CosyVoice2 generates five candidates, each candidate is independently scored by the AudioLLM for target-emotion fidelity and overall quality, and the highest-scoring candidate is retained.

\begin{table}[h]
\centering
\small
\setlength{\tabcolsep}{4.5pt}
\begin{tabular*}{\columnwidth}{l @{\extracolsep{\fill}} ccc}
\toprule
\textbf{System} & \textbf{WER~$\downarrow$} & \textbf{SIM~$\uparrow$} & \textbf{UTMOS~$\uparrow$} \\
\midrule
CosyVoice2       & 3.69   & 0.70   & 3.33   \\
EmoVoice          & 3.65   & 0.65   & 3.31   \\
\ours{} (Stage 1)          & 2.94   & 0.74   & 3.76  \\
CosyVoice2 re-generation   & 2.90   & 0.74   & 3.81  \\
EmoVoice re-generation     & 2.81   & 0.72   & 3.79  \\
Judge-reranked best-of-5   & 2.90   & 0.74   & 3.82  \\
\ours{} (Full)          & 2.73   & 0.73   & 3.93   \\
\bottomrule
\end{tabular*}
\caption{Objective evaluation on 3{,}000 emotional synthesis utterances. WER is computed via Whisper-large-v3-turbo; SIM is measured against the neutral speaker reference; UTMOS estimates perceptual quality.}
\label{tab:objective_eval}
\end{table}

All rows use the same 3{,}000 requests and evaluation metrics. The full \ours{} pipeline obtains the lowest WER and highest UTMOS while retaining comparable SIM. Judge-reranked best-of-5 provides little improvement over a single CosyVoice2 retry and remains below \ours{} in WER and UTMOS, while requiring five generations and five AudioLLM scoring calls per request. In the refinement stage of \ours{}, each flagged request instead requires one AudioLLM call to produce its refine instruction and one Refiner generation.

To make the scaled-set cascade more transparent, we also report stage-level accounting in Table~\ref{tab:large_cascade_accounting}. This separates utterances that pass the initial coarse filter, those recovered by Stage~1 re-generation, those flagged by Stage~2, and the remaining flagged subset after one refinement round.

\begin{table}[h]
\centering
\small
\begin{tabular*}{\columnwidth}{l @{\extracolsep{\fill}} rr}
\toprule
\textbf{Stage} & \textbf{\# Utt.} & \textbf{Rate} \\
\midrule
Total & 3{,}000 & 100.00\% \\
Initial pass & 2{,}367 & 78.90\% \\
After regen. & 2{,}753 & 91.76\% \\
Stage~2 flagged & 237 & 7.90\% \\
After 1st refine & 103 & 3.43\% \\
\bottomrule
\end{tabular*}
\caption{Stage-level accounting for the large-scale \ours{} cascade.}
\label{tab:large_cascade_accounting}
\end{table}

\section{Data Statistics}
\label{app:data_stats}

Table~\ref{tab:data_stats} summarizes the composition of \textsc{Refiner-DB}. The dataset contains ${\sim}$42K contrastive tuples sourced from six publicly available corpora, all used for training. In addition, we construct three non-overlapping evaluation splits from the same source corpora: 1{,}796 tuples for validation and 500 tuples for testing. There is no overlap in target recordings or text utterance instances among the training, validation, and test sets. We do not claim an unseen-speaker split for all corpora; same-speaker neutral prompts are part of the zero-shot cloning setup.

ESD \cite{zhou2021seen}, RAVDESS \cite{livingstone2018ryerson}, SAVEE\footnote{\url{http://kahlan.eps.surrey.ac.uk/savee}}, MESS \cite{morgan2019categorical}, and EmoVoice-DB \cite{yang2025emovoice} provide emotionally expressive recordings, while LibriTTS \cite{zen2019libritts} contributes read speech with natural prosodic variation. Where available, utterance-level emotion, speed, and pitch metadata are sourced from VccmDataset \cite{ji2025controlspeech}. For each target recording $y_\text{tgt}$, we synthesize a neutral counterpart $y_\text{neu}$ with CosyVoice2 \cite{du2024cosyvoice2} in zero-shot mode using a neutral prompt from the same speaker, and retain pairs only when the neutral synthesis satisfies WER${=}0$ and UTMOS${>}3$. The paired utterances therefore share text and speaker prompt, making prosody the dominant contrast while not eliminating all acoustic differences.

\begin{table}[h]
\centering
\small
\begin{tabular*}{\columnwidth}{l @{\extracolsep{\fill}} rr}
\toprule
\textbf{Corpus} & \textbf{\# Tuples} & \textbf{Type} \\
\midrule
ESD              & 14{,}000 & Emotional \\
RAVDESS          & 1{,}440  & Emotional \\
SAVEE            & 480      & Emotional \\
MESS             & 6{,}800  & Emotional \\
EmoVoice-DB      & 17{,}280 & Emotional \\
LibriTTS         & 2{,}000  & Regular \\
\midrule
\textbf{Total}   & ${\sim}$42{,}000 & --- \\
\bottomrule
\end{tabular*}
\caption{\textsc{Refiner-DB} composition by source corpus.}
\label{tab:data_stats}
\end{table}

The 500 test tuples are partitioned into three non-overlapping subsets: 200 neutral--emotional pairs for AudioLLM annotation capability assessment (Section~\ref{ssec:validation_annotation}), 100 TTS-generated utterances for AudioLLM diagnostic capability assessment (Section~\ref{ssec:validation_diagnostic}), and 200 tuples for instruction-following comparison (Section~\ref{ssec:main_results}).

\section{Pipeline Evaluation Setup}
\label{app:pipeline_setup}

\paragraph{Data preparation.} Using speakers from the evaluation set, we pair their neutral reference audio with text transcripts as speaker prompts for CosyVoice2 zero-shot generation, synthesizing 2{,}000 utterances per condition (\textit{Neutral} and \textit{Emotional}). For the Emotional condition, the target emotion label is provided by the dataset condition and used for all systems. Stage~1 performs up to two additional re-generation attempts after the initial generation, stopping early once WER${\leq}3.0\%$ and UTMOS${>}3.0$ are both satisfied; samples still failing after these attempts do not enter Stage~2. After Stage~1 filtering and Stage~2 AudioLLM diagnosis (score 1--10), utterances scoring ${<}\,5$ are flagged. We sample 100 flagged utterances per condition (200 total) for MOS-style human evaluation, directly testing each method's ability to recover from diagnosed defects. For \ours{}-Q, the same evaluation protocol and Refiner are used, but Qwen3-Omni Thinking replaces Gemini-3-pro for the complete inference-time Judge step, including diagnosis and instruction generation.

\paragraph{Condition details.} Table~\ref{tab:pipeline_eval} reports eight conditions in three groups:

\begin{itemize}
    \item \textbf{Before refinement} (Baseline): original flagged utterances, serving as the lower bound.

    \item \textbf{CosyVoice2 re-generation} (Re-generation, open-loop): re-synthesized from scratch by CosyVoice2 with a different random seed but the same text $T$, speaker prompt, and emotion instruction. This is the default untargeted retry strategy, using the same Stage~2 Judge trigger and one additional synthesis call as one-round \ours{}.

    \item \textbf{EmoVoice re-generation} (Re-generation, open-loop): same setup as above but using EmoVoice as the generator, testing cross-model recovery under the same one-correction budget.

    \item \textbf{\ours{}, Global-only} (Closed-loop): the Refiner retains the Judge-derived global emotion, speed, and pitch controls but receives no word-level stress or pause cues. All other inputs and evaluation settings match one-round \ours{}.

    \item \textbf{\ours{}, 1-round refinement} (Closed-loop): the flagged $y_\text{init}$ is diagnosed by Gemini-3-pro to produce instruction $I$; the Refiner takes $(y_\text{init}, I, T)$ and outputs corrected speech $y_R$.

    \item \textbf{\ours{}-Q, 1-round refinement} (Closed-loop): same as \ours{}, except that Qwen3-Omni Thinking performs the complete Stage~2 Judge step, including scoring, diagnosis, and instruction generation. The Refiner is unchanged and is not retrained.

    \item \textbf{\ours{}, 2-round refinement} (Closed-loop): after the first round produces $y_R^{(1)}$, it is re-diagnosed; if still ${<}\,5$, a second instruction $I^{(2)}$ is generated and $y_R^{(2)} = \text{Refiner}(y_R^{(1)}, I^{(2)}, T)$. No second round if the first already scores ${\geq}\,5$.

    \item \textbf{Human-instruction} (Closed-loop): a native English speaker with phonetics training first writes the refine instruction in the same structured format (global style + word-level stress/pause), replacing the AudioLLM output. Two additional professional annotators then review and revise the stress/pause choices until consensus. This isolates instruction quality and serves as a soft upper bound.
\end{itemize}

\paragraph{Evaluation protocol.} SIM is computed against the neutral speaker reference. Ten professional paid evaluators score each sample on MOS (1--5) in a blind test with all conditions shuffled and system identities hidden. Conditions are compared on matched texts, speakers, and target styles; evaluators may therefore hear the same text across systems, but presentation order is randomized. For paired preference evaluation, each comparison uses 30 flagged samples and the same 10 evaluators in a blind A/B setting. The order of systems is randomized per sample, and we report the aggregate preference rate as well as the number of samples for which at least 7/10 or 9/10 evaluators agree on the preferred system.

\section{Additional Robustness and Reliability Analyses}
\label{app:additional_analyses}

\subsection{Audio Conditioning Ablation}
\label{app:audio_conditioning_ablation}

The Refiner conditions on the initial utterance $y_\text{init}$, which provides an acoustic reference for preserving speaker identity and unmodified prosodic attributes. Removing this input reduces the model to instruction-conditioned synthesis from the text and speaker prompt, testing whether targeted correction benefits from explicitly seeing the utterance to be refined. This ablation is also a check on whether the system is genuinely using the defective input or mainly acting as a standalone expressive TTS model.

\begin{table}[h]
\centering
\small
\setlength{\tabcolsep}{4.5pt}
\begin{tabular*}{\columnwidth}{l @{\extracolsep{\fill}} cccc}
\toprule
\textbf{Variant} & \textbf{WER}$\downarrow$ & \textbf{SIM}$\uparrow$ & \textbf{MOS}$\uparrow$ & \textbf{Avg.\ MOS-I}$\uparrow$ \\
\midrule
Refiner & 2.90 & 0.68 & 4.21$_{\pm.14}$ & 4.21$_{\pm.09}$ \\
w/o $y_\text{init}$ & 2.91 & 0.62 & 4.17$_{\pm.12}$ & 3.94$_{\pm.11}$ \\
\bottomrule
\end{tabular*}
\caption{Ablation of conditioning on the initial utterance.}
\label{tab:ablation_audio_conditioning}
\end{table}

Removing $y_\text{init}$ only slightly decreases MOS from 4.21 to 4.17, showing that the model can still generate broadly natural speech from the text, speaker prompt, and instruction. However, SIM drops from 0.68 to 0.62 and Avg.\ MOS-I decreases from 4.21 to 3.94. This indicates that the initial utterance is important for targeted refinement: without this acoustic reference, the model is more likely to drift from the intended speaker/prosodic identity and, in some instruction-following cases, change prosodic regions that were not requested. The full Refiner therefore better balances naturalness, preservation, and instruction execution.

\subsection{Counterfactual Instruction Robustness}
\label{app:counterfactual}

To test whether the Refiner blindly executes erroneous local edits, we construct a counterfactual set of 40 utterances containing linguistically inappropriate stress placements on function words or pause placements between tightly coupled constituents. For each utterance, one correct stress/pause position is replaced by a counterfactual one, allowing us to check both wrong-position execution and correct-position preservation.

\begin{table}[h]
\centering
\small
\begin{tabular*}{\columnwidth}{l @{\extracolsep{\fill}} cc}
\toprule
\textbf{Setting} & \textbf{WER}$\downarrow$ & \textbf{SIM}$\uparrow$ \\
\midrule
Before refinement & 1.92 & 0.68 \\
With GT instruction & 1.72 & 0.67 \\
With counterfactual instruction & 1.87 & 0.68 \\
\bottomrule
\end{tabular*}
\caption{Objective results under counterfactual local stress/pause instructions.}
\label{tab:counterfactual_objective}
\end{table}

The counterfactual instruction affects refinement, but WER still improves over the pre-refinement audio. This suggests that the Refiner does not simply sacrifice content fidelity to execute a noisy instruction. One reason is the training setup: during Refiner training, instructions are paired with correct target audio, so the model mainly learns acoustically plausible corrections that move the input toward a better target realization, rather than arbitrary local command execution. This test is limited to a small counterfactual set and does not fully measure false-edit rates on natural unrequested positions.

\begin{table}[h]
\centering
\small
\begin{tabular*}{\columnwidth}{l @{\extracolsep{\fill}} cc}
\toprule
\textbf{Type} & \textbf{Wrong Exe. Rate} & \textbf{Corr. Pres. Rate} \\
\midrule
Stress & 10\% & 34\% \\
Pause & 28\% & 52\% \\
\bottomrule
\end{tabular*}
\caption{Professional annotation of counterfactual instruction execution. Exe. and Pres. denote execution and preservation.}
\label{tab:counterfactual_human}
\end{table}

These results suggest partial robustness to over-diagnosed local edits: erroneous local edits are usually not introduced. The non-zero preservation rates should be interpreted more cautiously; they show that some emotionally salient stress/pause cues can remain natural even when omitted from the instruction, but they do not eliminate the risk of under-diagnosis.

\subsection{Iteration Budget}
\label{app:iteration_budget}

We further check a 3-round setting to evaluate whether additional iterations continue to help.

\begin{table}[h]
\centering
\small
\begin{tabular*}{\columnwidth}{l @{\extracolsep{\fill}} cc cc}
\toprule
\multirow[c]{2}{*}{\textbf{Setting}} & \multicolumn{2}{c}{\textbf{Neutral}} & \multicolumn{2}{c}{\textbf{Emotional}} \\
\cmidrule(lr){2-3} \cmidrule(lr){4-5}
 & \textbf{WER}$\downarrow$ & \textbf{SIM}$\uparrow$ & \textbf{WER}$\downarrow$ & \textbf{SIM}$\uparrow$ \\
\midrule
1-round & 1.86 & 0.66 & 1.79 & 0.65 \\
2-round & 1.85 & 0.64 & 1.80 & 0.61 \\
3-round & 1.85 & 0.60 & 1.80 & 0.58 \\
\bottomrule
\end{tabular*}
\caption{Objective results with additional refinement rounds.}
\label{tab:three_round}
\end{table}

The objective gains saturate after one to two rounds, while SIM gradually decreases as later rounds refine previously generated audio rather than the original audio. AudioLLM over- or under-diagnosis may also contribute to this accumulation through unnecessary edits or residual defects. We therefore view one or two rounds as the practical operating point.

\subsection{Human Evaluation Reliability}
\label{app:human_reliability}

For MOS-style ratings, we compute Intraclass Correlation Coefficient ICC(2,k), which measures the reliability of averaged continuous scores from multiple raters. Following Cicchetti's guideline \cite{cicchetti1994guidelines}, values from 0.60 to 0.74 are considered good and 0.75 to 1.00 excellent.

\begin{table*}[t]
\centering
\small
\begin{tabular*}{\textwidth}{l @{\extracolsep{\fill}} cccccc}
\toprule
\textbf{System} & \textbf{MOS} & \textbf{Emotion} & \textbf{Pause} & \textbf{Stress} & \textbf{Pitch} & \textbf{Speed} \\
\midrule
CosyVoice2 & 0.784 & 0.562 & 0.915 & 0.496 & 0.728 & 0.848 \\
CosyVoice2-marker & 0.549 & 0.588 & 0.943 & 0.598 & 0.729 & 0.865 \\
EmoVoice & 0.496 & 0.706 & 0.899 & 0.694 & 0.683 & 0.823 \\
Refiner & 0.791 & 0.679 & 0.886 & 0.677 & 0.744 & 0.835 \\
\bottomrule
\end{tabular*}
\caption{ICC(2,k) reliability for MOS and MOS-I ratings.}
\label{tab:icc_reliability}
\end{table*}

Most MOS/MOS-I reliability values are in the good range or above, and the Refiner shows good consistency on the prosody-related dimensions. Some dimensions remain below the good range, reflecting the inherent difficulty of fine-grained subjective prosody evaluation.

\subsection{Speaker Similarity Trade-off}
\label{app:speaker_similarity_tradeoff}

SIM in our main experiments is computed against a neutral speaker reference. This metric can decrease when emotional prosody moves away from the neutral reference and therefore conflates intended emotional variation with speaker-identity drift. To examine speaker preservation under emotional generation, we additionally compute same-speaker expressive-reference similarity (SIM-EXP). For each generated output $x$, we average its similarity to five different-utterance references $r_i$ from the same speaker and target emotion:

\begin{equation}
\mathrm{SIM\text{-}EXP}(x) = \frac{1}{5}\sum_{i=1}^{5}
\cos\!\left(e(x), e(r_i)\right),
\label{eq:sim_exp}
\end{equation}
where $e(\cdot)$ denotes the normalized WavLM speaker embedding.

\begin{table}[h]
\centering
\small
\begin{tabular*}{\columnwidth}{l @{\extracolsep{\fill}} cc}
\toprule
\textbf{System} & \textbf{Neutral SIM}$\uparrow$ & \textbf{SIM-EXP}$\uparrow$ \\
\midrule
CosyVoice2 & \textbf{0.72} & 0.65 \\
EmoVoice & 0.67 & 0.75 \\
Refiner & 0.68 & \textbf{0.77} \\
\bottomrule
\end{tabular*}
\caption{Speaker similarity against a neutral reference and five same-speaker, same-emotion expressive references.}
\label{tab:sim_exp}
\end{table}

CosyVoice2 has the highest neutral-reference SIM but the lowest SIM-EXP, consistent with its weaker target-emotion realization in Table~\ref{tab:main_results} (Emotion MOS-I: 2.25). The Refiner retains neutral-reference SIM comparable to EmoVoice while achieving the highest SIM-EXP. These results suggest that the SIM reduction under emotional generation primarily reflects intended emotional variation rather than substantial speaker drift. Similar patterns have been reported in expressive TTS: CoCoEmo observes that speaker similarity can decrease as emotional similarity increases \cite{cocoemo2026}, and DiEmo-TTS shows that speaker embeddings can form emotion-dependent sub-clusters \cite{diemo_tts2025}.

\subsection{Construction Cost Estimate}
\label{app:construction_cost}

For \textsc{Refiner-DB} construction, each tuple requires one AudioLLM call over two audio clips. Including audio, transcript, and prompt, each tuple uses about 2.7K input tokens and 0.6K output tokens. For 42K tuples, this corresponds to about 113.4M input tokens and 25.2M output tokens. Using Gemini-3-pro pricing of \$4 per 1M input tokens and \$18 per 1M output tokens, the estimated construction cost is about \$907.2. At inference time, each retained utterance requires one Judge call, and each flagged utterance additionally requires one Refiner generation per refinement round; this is why our intended deployment setting is offline batch synthesis.

\subsection{Prompt-only TTS Baselines}
\label{app:prompt_only}

We also conducted small-sample listening trials with Qwen3-TTS \cite{qwen3tts2026} and MOSS-TTS \cite{mosstts}. In these trials, the models did not reliably execute fine-grained word-level stress or pause edits through prompts alone, making them difficult to use as comparable refiners for this task. This observation is preliminary rather than a full MOS-style baseline, and we do not claim that all future prompt-only or SSML-like systems would fail. The key distinction is that prompt-only optimization is limited by the controllability exposed by the original TTS model, whereas \ours{} performs correction through a separate post-generation Refiner.

\section{Instruction-Following Baselines}
\label{app:baselines}

\textbf{CosyVoice2} \cite{du2024cosyvoice2} is a streaming TTS that accepts natural-language instructions alongside target text and generates speech from scratch. It must interpret open-ended instructions, which often leads to verbatim reading of the instruction rather than executing it as a prosodic directive.

\textbf{CosyVoice2-marker} augments CosyVoice2 with explicit prosodic markup tags (e.g., \texttt{<strong>}, \texttt{<breath>}) instead of free-form language. This provides unambiguous structural directives and avoids verbatim reading, but can cause mispronunciation at modified positions. We tune prompts on the validation set and use the best-performing markup variant for the comparison.

\textbf{EmoVoice} \cite{yang2025emovoice} is the pretrained 1.5B backbone \textit{without} refinement fine-tuning, isolating the contribution of our training procedure from the backbone's inherent capabilities.

Speech editing methods (VoiceCraft \cite{peng2024voicecraft}, SpeechX \cite{wang2024speechx}) are excluded as they address content-level word replacement and cannot accept prosodic instructions.

\section{Implementation Details}
\label{app:implementation}

\paragraph{Model architecture.}
The Refiner is built on EmoVoice-1.5B, which uses Qwen2.5-1.5B (hidden dimension 896) as the language model backbone. Audio is represented as 3-layer grouped semantic codes from the CosyVoice codec (audio vocabulary size 4{,}160 per layer). We initialize from the pretrained EmoVoice checkpoint and fine-tune all LLM parameters (encoder frozen).

\paragraph{Training setup.}
We fine-tune with AdamW (learning rate $5{\times}10^{-6}$, no weight decay) using a warmup schedule (1{,}000 warmup steps, 300K total steps) and a batch size of 8.

\paragraph{Hyperparameters.}
The position-weighting hyperparameters are $\alpha{=}0.2$ (stress) and $\beta{=}0.3$ (pause) in Equation~\ref{eq:weight}. The structural alignment loss weight is $\lambda{=}0.1$ in Equation~\ref{eq:total_loss}. These values were selected based on validation-set MOS-I and WER trends without extensive search.

\paragraph{AudioLLM versions.}
We accessed Gemini-3-pro through Google's official Gemini API using the model code \texttt{gemini-3-pro-preview} in January 2026. For Qwen3-Omni Thinking, we use the official \texttt{Qwen/Qwen3-Omni-30B-A3B-Thinking} checkpoint.

\paragraph{Inference.}
The Judge score threshold is $\tau_\text{judge}{=}5$. Stage~1 filtering uses Whisper-large-v3-turbo for WER computation and UTMOS for quality estimation, with WER${\leq}3.0\%$ and UTMOS${>}3.0$ as the pass condition. The Refiner uses deterministic greedy decoding (argmax, without sampling) with a maximum of 256 new tokens per utterance. AudioLLM responses are parsed as strict JSON according to the schemas in Appendix~\ref{app:prompts}; malformed outputs are retried once, and unresolved failures are excluded during data construction or treated as no-refinement fallbacks during inference. Because proprietary AudioLLM endpoints can change, we will release the prompts, raw parsed outputs, and per-call metadata where permitted.

\section{AudioLLM Evaluation Setup}
\label{app:eval_setup}

\paragraph{Data source.} Evaluation data is drawn from the 500-tuple test set held out from \textsc{Refiner-DB}. We use 200 neutral--expressive pairs for annotation evaluation and 100 TTS-generated utterances for diagnostic evaluation. For each emotional utterance, a neutral counterpart is synthesized via CosyVoice2 zero-shot generation, yielding 200 neutral--emotional pairs.

\paragraph{Diagnostic Capability (Section~\ref{ssec:validation_diagnostic}).} We select 50 Regular and 50 Emotional TTS-generated utterances with noticeable prosodic defects. Each utterance is presented to five AudioLLMs along with its text transcript; the AudioLLM diagnoses prosodic defects and generates refine instructions. Six non-technical evaluators with fluent English listening and reading proficiency independently listen to each utterance, review the AudioLLM's diagnosis, and rate its alignment with their own judgment on a 1--4 scale (4: fully aligned, 3: mostly aligned, 2: partially aligned, 1: not aligned). Scores are averaged across evaluators.

\paragraph{Annotation Capability (Section~\ref{ssec:validation_annotation}).} All 200 neutral--expressive pairs are used. Professional annotators with expertise in prosodic analysis listen to each pair and label the most prominent stressed words and pause positions that distinguish the expressive utterance from its neutral counterpart (see the annotation interface in Appendix~\ref{app:annotation_ui}). Each AudioLLM is independently prompted to annotate the same pairs, and its outputs are evaluated against the professional labels using micro recall and F1.

\section{Scoring Rubrics}
\label{app:scoring}

Evaluators receive the following rubrics before the blind test. Each sample is rated independently on MOS and five MOS-I dimensions. In Table~\ref{tab:pipeline_eval}, MOS measures whether the recovered utterance sounds natural after correction. In Table~\ref{tab:main_results}, MOS-I separately measures whether the specified instruction is executed on the intended control dimension, rather than overall naturalness. For local controls, MOS-I is our human subjective proxy for requested stress/pause execution and unintended changes that make the instruction less correctly or naturally realized.

\paragraph{MOS: Naturalness \& Audio Quality.} Rate how natural and human-like the speech sounds, regardless of whether instructions are followed. \textbf{5 (Excellent):} Completely natural, human-like, no artifacts. \textbf{4 (Good):} Natural, but with very minor artifacts. \textbf{3 (Fair):} Understandable, but somewhat robotic or buzzy. \textbf{2 (Poor):} Unnatural, distinct metallic or robotic sounds. \textbf{1 (Bad):} Completely unnatural, unintelligible artifacts.

\paragraph{MOS-I: Instruction Adherence.} Each of the five sub-dimensions (emotion, pitch, speed, stress, pause) is scored independently. Focus only on whether the specific instruction for that dimension is followed. \textbf{5 (Natural adherence):} The instruction is fully followed, and the result sounds natural and effortless. \textbf{4 (Followed, slightly unnatural):} The instruction is followed, but the execution feels slightly exaggerated or mechanical. \textbf{3 (Partially followed):} The instruction is partially followed; some aspects are correct, others are missing or weak. \textbf{2 (Barely followed):} The instruction is barely perceptible; very weak or mostly missing. \textbf{1 (Not followed):} The instruction is completely ignored or contradicted.

\section{AudioLLM Prompts}
\label{app:prompts}

This section provides the full prompts used for AudioLLM-based prosodic annotation (Section~\ref{ssec:data_pipeline}) and inference-time diagnosis (Section~\ref{ssec:pipeline}).

\subsection{Annotation Prompt (Training-Time)}
\label{app:annotation_prompt}

The following prompt is used during \textsc{Refiner-DB} construction (Section~\ref{ssec:data_pipeline}, Step~2). The AudioLLM receives Audio~A ($y_\text{neu}$, the neutral synthesis) and Audio~B ($y_\text{tgt}$, the expressive human recording) along with the shared transcript, and outputs a structured JSON specifying word-level prosodic differences. Global style attributes (emotion, speed, pitch) are extracted separately from dataset metadata and merged with the JSON output in post-processing to form the final refine instruction~$I$.

\begin{tcolorbox}[
  colback=gray!5,
  colframe=gray!60,
  fonttitle=\bfseries\small,
  title=Annotation Prompt,
  breakable,
  fontupper=\scriptsize\ttfamily,
  left=4pt, right=4pt, top=4pt, bottom=4pt
]
\textbf{SYSTEM INSTRUCTIONS}\\
You are a Prosody Comparison Expert. Your goal is to compare two audio recordings of the same text and identify word-level prosodic differences.\\[6pt]
\textbf{TASK}\\
1. Listen to both audios: Audio A is a neutral, prosodically flat synthesis. Audio B is an expressive human recording of the same text.\\
2. Compare A vs B: Identify where A fails to match B's prosody, focusing on two dimensions:\\
\hspace*{12pt}- stress: Words that are emphasized/stressed in B but not in A.\\
\hspace*{12pt}- pause: Positions where B has a noticeable pause but A does not (report the word before the pause).\\
3. Select top changes: Pick the 3--5 most salient prosodic differences.\\
4. Generate JSON: Output the result in strict JSON format.\\[6pt]
\textbf{OUTPUT FORMAT}\\
\{\\
\hspace*{12pt}"local\_edits": [\\
\hspace*{24pt}\{ "word": "<word>", "type": "stress" | "pause" \}\\
\hspace*{12pt}]\\
\}\\[6pt]
\textbf{CONSTRAINTS}\\
1. Type values: MUST be one of "stress" or "pause".\\
2. Word matching: The "word" field MUST be an exact word from the provided transcript.\\
3. Evidence-based: Only report differences you can clearly hear between A and B. Do not guess.\\
4. Output: RAW JSON ONLY. No markdown blocks, no explanation.
\end{tcolorbox}

\subsection{Diagnostic Prompt (Inference-Time)}
\label{app:diagnostic_prompt}

The following prompt is used during Stage~2 inference-time diagnosis (Section~\ref{ssec:pipeline}). Unlike the annotation prompt, which compares two audio recordings, the diagnostic prompt receives a single TTS-generated utterance $y_\text{init}$ along with its transcript and, when available, the target emotion label from the synthesis condition. The AudioLLM first scores prosodic naturalness and emotional fidelity (1--10). If the score is 5 or above, only the score is returned and the utterance is accepted. Otherwise, the AudioLLM produces a full diagnosis covering global style (emotion, speed, pitch) and local prosody (word-level stress and pause positions), along with a natural-language refine instruction forwarded to Stage~3.

\begin{tcolorbox}[
  colback=gray!5,
  colframe=gray!60,
  fonttitle=\bfseries\small,
  title=Diagnostic Prompt (Inference-Time),
  breakable,
  fontupper=\scriptsize\ttfamily,
  left=4pt, right=4pt, top=4pt, bottom=4pt
]
\textbf{SYSTEM INSTRUCTIONS}\\
You are a Prosodic Quality Judge. Your goal is to diagnose prosodic defects in a TTS-generated utterance and, if needed, produce a structured refine instruction for correction.\\[6pt]
\textbf{TASK}\\
1. Listen to the audio and read the transcript carefully.\\
2. Score: Rate overall prosodic naturalness and emotional fidelity on a 1--10 scale (10: perfect, 1: completely unnatural).\\
3. If score >= 5: Output only the score. No further diagnosis is needed.\\
4. If score < 5: Produce a full diagnosis:\\
\hspace*{12pt}a. Diagnose global style: Specify the target emotion, speed, and pitch using the strict enums below.\\
\hspace*{12pt}b. Diagnose local prosody: Identify the 3--5 most salient word-level prosodic defects, focusing on two dimensions:\\
\hspace*{24pt}- stress: Words that should be emphasized but are not.\\
\hspace*{24pt}- pause: Positions where a pause is missing (report the word before the pause position).\\
\hspace*{12pt}c. Generate a natural-language refine instruction summarizing all changes.\\
5. Generate JSON: Output the result in strict JSON format.\\[6pt]
\textbf{GLOBAL STYLE ENUMS (STRICT)}\\
\begin{tabular}{@{}l@{\hspace{6pt}}l@{}}
emotion: & ["Angry", "Contempt", "Disgusted", "Fear",\\ &"Happy", "Sad", "Surprised", "Neutral"] \\
speed:   & ["Slow", "Moderate", "Fast"] \\
pitch:   & ["Low", "Moderate", "High"] \\
\end{tabular}\\[6pt]
\textbf{OUTPUT FORMAT}\\
If score >= 5:\\
\{ "score": <1-10> \}\\[4pt]
If score < 5:\\
\{\\
\hspace*{12pt}"score": <1-10>,\\
\hspace*{12pt}"global\_style": \{\\
\hspace*{24pt}"emotion": "<target>",\\
\hspace*{24pt}"speed": "<target>",\\
\hspace*{24pt}"pitch": "<target>"\\
\hspace*{12pt}\},\\
\hspace*{12pt}"local\_edits": [\\
\hspace*{24pt}\{ "word": "<word>", "type": "stress" | "pause" \}\\
\hspace*{12pt}],\\
\hspace*{12pt}"instruction": "<natural-language refine instruction>"\\
\}\\[6pt]
\textbf{CONSTRAINTS}\\
1. Strict enums: Global style values MUST be chosen from the provided lists.\\
2. Word matching: The "word" field MUST be an exact word from the provided transcript.\\
3. Type values: MUST be one of "stress" or "pause".\\
4. Instruction: Summarize all suggested changes (global and local) in one concise natural-language sentence.\\
5. Output: RAW JSON ONLY. No markdown blocks, no explanation.
\end{tcolorbox}

\begin{figure*}
    \centering
    \includegraphics[width=\textwidth]{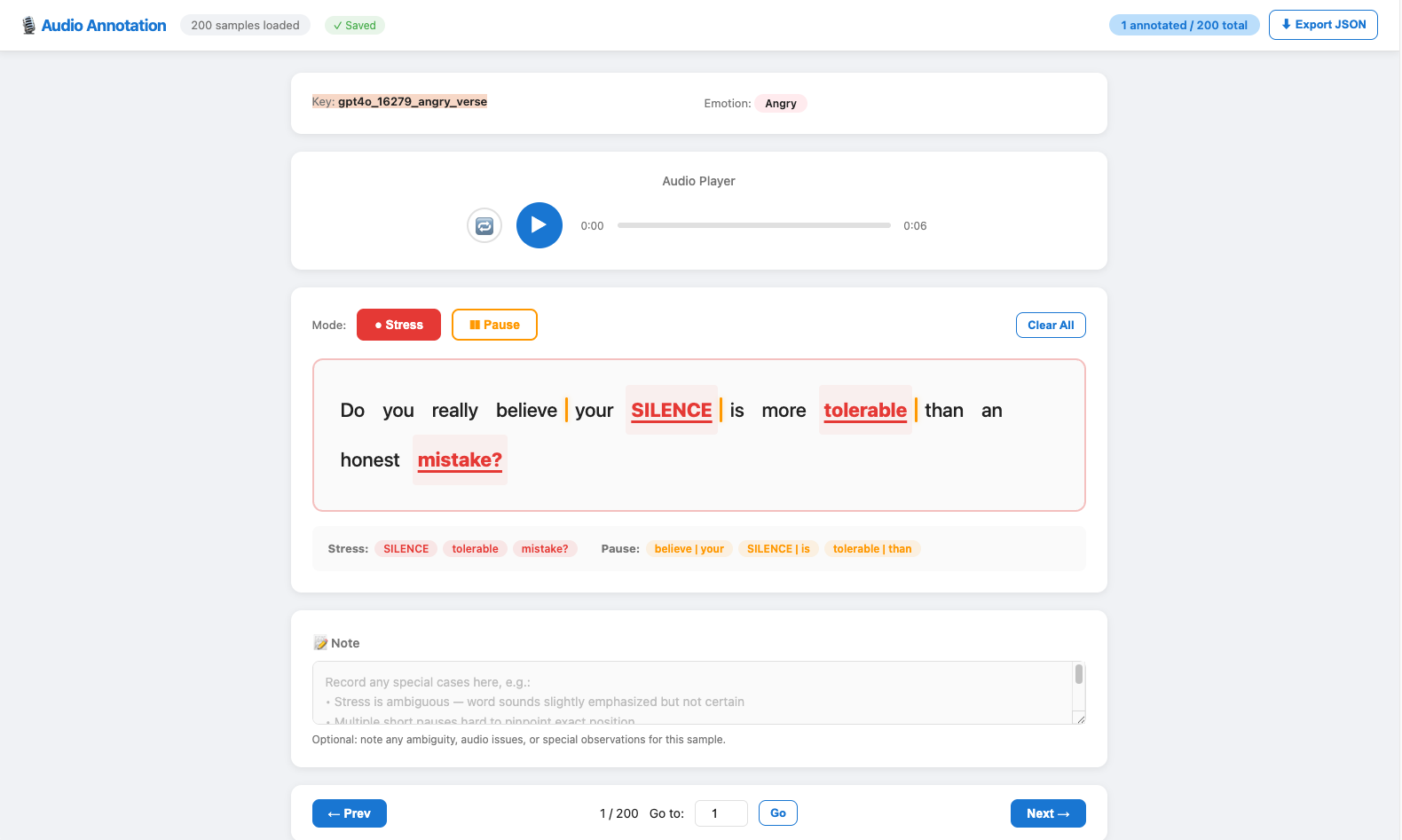}
    \caption{Screenshot of the web-based human annotation interface used for collecting prosodic ground-truth labels (Section~\ref{ssec:validation_annotation}). Annotators toggle between \textit{Stress} and \textit{Pause} modes, clicking words or inter-word gaps to mark stressed words (red) and pause positions (vertical bars). A live preview and free-text note field are provided below.}
    \label{fig:annotation_ui}
\end{figure*}
\newpage
\section{Human Annotation Interface}
\label{app:annotation_ui}

To collect professional human labels for evaluating AudioLLM annotation quality (Section~\ref{ssec:validation_annotation}), we build a web-based annotation tool (Figure~\ref{fig:annotation_ui}). For each of the 200 neutral--expressive utterance pairs, annotators listen to the expressive recording and interact with the displayed transcript in two modes:

\begin{itemize}
    \item \textbf{Stress mode}: clicking a word toggles it as a stressed word (highlighted in red).
    \item \textbf{Pause mode}: clicking the gap between two words toggles a pause marker at that position (indicated by a vertical bar).
\end{itemize}

Selected stress words and pause positions are displayed in a live preview below the text. Annotators can also leave free-text notes for ambiguous cases. Annotations are auto-saved and exportable as structured JSON, with each entry recording the list of stressed word indices and pause slot indices for downstream evaluation.

\end{document}